\documentclass[12pt,a4paper,titlepage]{article}

\usepackage{amsmath}
\usepackage{amsfonts}
\usepackage{amssymb}
\usepackage[pdftex]{graphicx}

\begin{document}

\begin{titlepage}
\begin{center}
\begin{Huge} Athena: \end{Huge} \\ 
\begin{LARGE} Modular CAM/CAD Software for Synthetic Biology \end{LARGE} 

\end{center}

Authors: 

Deepak Chandran$^1$ \\
Frank T. Bergmann$^{1,2}$ \\
Herbert M. Sauro$^1$ 

\small
$^{1}$Department of Bioengineering, University of Washington, Box 355061,\\
William H. Foege Building, Room N210E, \\ 
Seattle, WA, USA 98195-5061 \\ 
$^2$ Keck Graduate Institute, \\ 
535 Watson Drive, Claremont, CA, USA, 91711

email: deepakc@u.washington.edu

keywords: synthetic biology, modularity

\end{titlepage}

\begin{abstract}
Synthetic biology is the engineering of cellular networks. It combines principles of engineering and the knowledge of biological networks to program the behavior of cells. Computational modeling techniques in conjunction with molecular biology techniques have been successful in constructing biological devices such as switches, oscillators, and gates. The ambition of synthetic biology is to construct complex systems from such fundamental devices, much in the same way electronic circuits are built from basic parts. As this ambition becomes a reality, engineering concepts such as interchangeable parts and encapsulation will find their way into biology. We realize that there is a need for computational tools that would support such engineering concepts in biology. As a solution, we have developed the software Athena that allows biological models to be constructed as modules. Modules can be connected to one another without altering the modules themselves. In addition, Athena houses various tools useful for designing synthetic networks including tools to perform simulations, automatically derive transcription rate expressions, and view and edit synthetic DNA sequences. New tools can be incorporated into Athena without modifying existing program via a plugin interface, IronPython scripts, Systems Biology Workbench interfacing and the R statistical language. The program is currently for Windows operating systems, and the source code for Athena is made freely available through CodePlex, www.codeplex.com/athena.

\end{abstract}

\section{Introduction}

Over the past few decades, biological protocols for inserting genes and manipulating a cell's innate behaviour have become common practice \cite{RecombDNA}. At the same time, systems biology has provided the ability to mathematically represent and analyze biological systems. The ability to computationally analyze a biological system and the procedures for constructing that system together have given birth to the emerging field of synthetic biology \cite{Weiss}. Synthetic biology brings the vision of constructing organisms with designed genetic and metabolic networks that perform functions that are of interest to the designer. To date, such synthetic networks can mimic functions such as logic gates, switches, and oscillators \cite{intro1,intro3,intro4,intro5,intro6,intro8,intro9,intro10}. These synthetic networks have been modelled and simulated to reproduce the behaviour seen in a cell, demonstrating the fact that it is possible to design, model, and build biological circuits in much the same way as electronic circuits.

Synthetic biology aims to build networks that are composable, so that one designed network can be reused in different settings, allowing large networks to be built from simpler ones \cite{Weiss}. This aim has been pioneered by the BioBricks Foundation \cite{parts,biobrick}, which has introduced various new concepts and terms to fuel synthetic biology as an engineering discipline. The BioBricks Foundation defined the notion of a BioBrick ``part", the basic unit used for construction of a synthetic network. Current BioBrick parts include promoters, protein coding regions, ribosome binding sites, transcriptional terminators, and other functional components which can be ligated together to construct regulatory networks. The MIT Registry of Parts \cite{registry} hosts a database of such parts. Just as electric engineers are able to purchase parts to build a circuit, the vision is to do the same for biological systems.

While the term ``part" refers to a physical entity, the term ``device" \cite{Weiss,popsdevices} is used to identify a functional entity. A synthetic biology device is a composition of various parts that performs a well defined function. Devices can be as simple as a regulatory region followed by a gene, or it can be an entire regulatory pathway, as long as it has a well defined function. The inspiration behind the concept of a device is that well behaved devices can be connected together in different ways to create complex systems easily and reliably. Various ongoing efforts hope to resolve issues such as proper definitions for parts and devices, because well defined behaviours and descriptions are needed for synthetic biology to move forward. One proposed interface for connecting two devices is through ``PoPS" inputs and outputs \cite{pops}. Polymerase Per Second, or PoPS, refers the rate at which RNA polymerase is moving through a given region of DNA, much like current through a wire. This notion is important for synthetic biology because it allows a part such as a protein coding region to define its transcription rate independent of which promoter lies upstream of it: the promoter provides the PoPS current as a function of transcription factor concentrations, and the gene receives that current and produces a proportional amount of mRNA transcripts. The PoPS current travels through the gene and to the next part. A transcriptional terminator stops the current. Since much of the present synthetic biology relies on rewiring regulatory networks by altering promoter-gene combinations, the PoPS terminology is a useful way of defining parts without referring to specific proteins. 

\textit{Available software tools}
The software presented in this work, called Athena, allows modular construction of models as well as supports various features, such as sequence information, that is needed for synthetic biology. In order to select other features that Athena should support, we began by analyzing popular existing software packages. For this we selected CellDesigner \cite{CellDesigner}, JDesigner \cite{Bergmann_sbw}, and BioTapestry \cite{biotapestry} as representative modeling environments of the systems biology community and BioJADE as representative visual modeling environments for synthetic biology. 

CellDesigner \cite{CellDesigner,Kitano2}, developed by the Systems Biology Institute, Tokyo, is a visual modeling environment, focusing on representing biochemical networks employing the process-diagram notation. This allows for a complex visual representation of all parts of biochemical modeling. CellDesigner also allows for time-course simulations and works with the Systems Biology Workbench, allowing a modeler, to quickly analyze the current model with other software tools. CellDesigner 4.0 also features a plug-in API allowing 3rd parties to extend its capabilities. 

JDesigner \cite{Bergmann_sbw} operates in a similar way to CellDesigner, with a stronger focus on analysis and simulation of the created model than the visual representation. JDesigner automatically assigns kinetic laws to created reactions, speeding up the kinetic modeling aspect. JDesigner also assists modelers by providing a list of inbuilt kinetic laws that is filtered to match the reaction topology the modeler created. By tight integration with the Systems Biology Workbench, JDesigner integrates time course simulation, steady state and metabolic control analysis as well as structural analysis of the network. JDesigner can also be extended through the Systems Biology Workbench.

Both CellDesigner and JDesigner use the Systems Biology Markup Language (SBML) \cite{SBML} as their native format, making it straight forward to use models created with over 120 other software tools (as seen on the sbml.org homepage). Both groups have also joined the Systems Biology Graphical Notation (SBGN) initiative \cite{SBGN}, thus enabling their software to exchange the visual description of the computational model.

BioTapestry \cite{biotapestry} is exclusively designed to build and visualize (large) gene-regulatory networks. Models can be created either by ``drawing" genes, similar to the CellDesigner and JDesigner approach, or by importing them from gene expression data. BioTapestry also features the concept of sub-models, where certain parts of a current model can be active or inactive at any given time. This feature is used extensively in developmental network models \cite{biotapestry2}.

BioJADE \cite{biojade} was probably the first visual software modeling tool for Synthetic Biology. It was developed closely with the BioBricks repositories and allows the design and simulation of synthetic biological systems. The BioBricks repositories take a central role in BioJADE. Existing parts can be accessed and new parts created in the repository. BioJADE employs techniques very similar to those used in electronic circuit designs, where new parts can be assembled from existing parts in a schematic view. BioJADE provides simulation capabilities relying on established stochastic simulation procedures. While BioJADE is open source, the connectivity to the central BioBricks repositories is currently not publicly available. As the only tool in this brief survey, BioJADE lacks SBML export and relies on direct database access or a proprietary XML format for model exchange. 

While currently existing tools such as JDesigner \cite{Bergmann_sbw, BioSpice} and CellDesigner \cite{CellDesigner} allow analysis of a single network, there are few such as BioJADE \cite{biojade} and BioTapestry \cite{biotapestry} that allow users to create a new network by connecting existing networks. However, synthetic biology requires models to be composable. To reach this goal, the hurdle is that the concept of ``module" is unclear for biological networks, particularly an engineerable module. We have developed the application Athena in order to demonstrate a working concept of a biological module and how this concept of a biological module can be used to construct synthetic modular networks. In addition, Athena contains various features and tools oriented toward synthetic biology such as access to a regulatory elements database, sequence editing, automatic transcription rate derivation, and interface with packages such as the Systems Biology Workbench (SBW) \cite{Hucka_SBW,Bergmann_sbw} and R \cite{R}.

\section{Approach}

The core components of Athena are designed for modular design of biological networks as well as to support various data that are needed for synthetic genetic networks, such as PoPS rates and sequence information. Additional functionalities are added to Athena through the plug-in interface, which is explained in section \ref{featuresSection}. 
Studying other software features allowed us to integrate the various positive attributes of each software into Athena, in addition to the modular design methodology. The following list summarizes the key attributes that we incorporated into Athena.

\begin{itemize}

\item Composable models: Larger models can be constructed using smaller models without altering the smaller models.
\item Support synthetic biology terminology: A common terminology used in current synthetic biology is the PoPS rate, which is incorporated into the genetic parts in Athena.
\item Database support: Access to a parts database that will allow users to build concrete designs. 
\item Standard formats: \- Supporting a standard format like SBML will allow Athena to
export models in order to analyze them with a growing variety of available software solutions. 
\item Familiar visual representations: Choosing visual representations that are familiar to 
the synthetic biology community will allow users to become quickly acquainted with the software.
\item Extensibility: Athena will allow 3rd parties to extend it using dedicated plug-ins, SBW Modules 
and custom scripts.
\end{itemize}

Athena is implemented in the C\# language using the .NET 2.0 framework.

\section{Methods}

\subsection{Network Construction}

Athena supports the construction of genetic, metabolic, or signalling networks. While signalling networks and metabolic networks are modelled identically to other visual applications such as JDesigner and CellDesigner, the genetic networks incorporate the notion of PoPS in order to support the current methodology in synthetic biology. Each genetic part, such as a promoter, gene, spacer DNA, or terminator, has a PoPS rate associated with it. The PoPS rate of a promoter is a function of transcription factor concentrations. The PoPS rate for other genetic parts are determined by the promoter(s) that lie upstream of those parts. When the user does not specify a function, Athena automatically assigns the PoPS rates when parts are connected to one another by looking at the PoPS of upstream parts. See Figure \ref{fig:01}

\subsection{Modularity in Athena}
A module is a composable model. A larger model can be constructed by connecting existing modules together. There are two fundamental ways by which modules can be connected in Athena.

The first manner of connecting two modules is by declaring shared molecular species between two or more modules. For example, suppose one person designs a module representing Glycolysis, and another designs a module for the Citric Acid Cycle. The manner in which these two modules will be joined in Athena is by declaring (visually) that the species called ``Pyruvate" in the Glycolysis module is the same as the ``Pyruvate" in the Citric Acid Cycle module. Athena does not modify the original modules when constructing the new combined network. A screenshot of this form of connecting is shown in Figure \ref{fig:02}. Athena allows for multiple connections like these to be made between two modules or between multiple modules.

Connecting two modules by declaring shared species can have various uses. One noteworthy use is that it allows for a user to merge existing SBML models that may share similar species, just as the modules in Figure \ref{fig:02} share the Pyruvate species. Two SBML files can be loaded into Athena and converted to modules; the species that are shared between the two models can be merged as shown in Figure \ref{fig:02}. This model containing the two connected modules can be exported as a single SBML file. The new SBML model would represent a merged form of the two other SBML files. Hence, Athena can be used to merge SBML files that contain the same species in them.

The second manner in which two modules can be connected is through PoPS. This form of connection only applies to modules with a genetic network, i.e. which contain a strand of DNA. Two such modules are connected by ligating their DNA strands together. The PoPS rate at the end of the first module will be carried over to the start of the second module, which allows information to pass from the first module to the second via PoPS rates. A classical example of such a module is the PoPS inverter, which is defined as a device that produces a low PoPS output for a high PoPS input (and vice versa). 

This device is composed of the coding region for a repressor protein (ribosome binding site is not drawn) and a promoter containing the operating site of that repressor. The production of the repressor is controlled by parts lying upstream of the inverter. If the incoming PoPS is high, then the repressor concentration will be high, which will prevent RNA polymerase from binding the promoter (labeled as p1 in Figure \ref{fig:03}, lowering its PoPS. Conversely, if the incoming PoPS is low, then the promoter will be active, and the PoPS will be high. This device can be placed upstream of another module, as shown in Figure \ref{fig:03}, allowing the PoPS output from the inverter to serve as the PoPS input to the next module. 

\subsubsection{Crosstalk Warnings}

Since biological modules generally contain proteins or other molecules that diffuse throughout the cell, if two identical modules are placed in the same cell, they will interfere with one another -- the proteins from one module will bind to the binding sites on the other module (and vice versa). This problem is called ``crosstalk" between modules. Athena detects crosstalk when two modules use the same species within the same compartment and signals a warning message that informs the users of the two species that are interfering with one another. 

\subsection{User Interface}

The graphical interface in Athena is intended to maximize the information displayed on the screen without sacrificing the visual appeal. This is achieved by having a tabular window (on the right in Figure \ref{fig:04} that shows all the information that is not shown in the graphical window. While the central canvas shows the connections and the basic architecture of a network, the tables on the side show information such as the rate of each reaction, the parameter values, the PoPS values, and concentrations of molecular species. The table can also be used to modify any of the values that are displayed in it, thus providing a table based interface for modifying a model. This table changes its content depending on the selected item. For instance, when different modules are selected, the table would show the kinetics pertaining to the selected module, independent of what is outside the module. When items other than modules, such as reactions, genetic parts, or species, are selected, then the table will show properties specific to those items. For genetic parts that can generate PoPS, such as promoters, the table provides the user with transcription rate expressions that mimic logic gates such as NOT, OR, AND, NOR, NAND, and XOR as well as simple activation and repression (all of these rates are derived using the automatic transcription rate derivation tool explained in Section \ref{transcriptionRateTool}.

Since Athena is designed for the modular design of networks, it allows a user to load more than one module onto the screen and analyze each separately as well the entire network with all the modules. Not only will the tabular view adjust its display depending on which module is selected, but the simulation tools will also simulate the selected module as an independent network. This allows a user to analyze and make modifications to individual modules without reloading each one.

It is anticipated that users would at times prefer to encapsulate a module (i.e. hide the inner components of a module) when building larger networks. To facilitate this perspective visually, Athena modules can be viewed in an encapsulated mode, where only the interfacing parts or species are visible. 

\label{featuresSection}
\subsection{Built-in Plugins}

Various tools are available for providing additional support. The database tool allows a user to search through all transcriptional regulations listed in the RegulonDB database \cite{RegulonDB} and automatically find transcription factors that bind to a particular promoter region. The sequence viewing tool allows a user to view the sequence of a selected strand of DNA and make changes to the sequence. The transcription rate tool provides the convenience for a user to derive the transcriptional rate formula for any combination of transcription factors. The script console tool allows users to interface with Athena via Python scripts. Each of these tools are described in more detail below.

\subsubsection{Database Tool}

The incentive behind database support is to allow an easy process by which a hypothetical network can be converted to a real network. For example, a user might construct a conceptual model of a bi-stable switch. After simulation and other forms of analysis, the design might be ready for construction. Athena allows the user to replace each promoter, gene, and protein in the network with items listed in RegulonDB, which houses sequences of 4579 genes, 1560 transcription factor binding sites, 1492 promoters, 179 ribosome binding sites, and 182 transcriptional terminators from \textit{E. coli}, which also includes the sequence information for each. More importantly, RegulonDB contains network information, a list of 2237 pairs of transcription factors and their target operator site. Athena utilizes the network information to find the matching transcription factor for any given operator site (or vice versa) in the hypothetical network. We chose to use RegulonDB because it contains a large amount of network information, such as which protein regulates which operator site, which is not present in other databases that house information about genes. Figure \ref{fig:05} illustrates the interface to the database plug-in.

\subsubsection{Sequence Viewing Tool}

The sequence tool allows users to view the entire DNA sequence composed of multiple parts as well as edit the sequences, as shown in Figure \ref{fig:06}. Including a sequence tool allows a user to construct a network, simulate or analyze it, and generate the final sequence for that network.

\label{transcriptionRateTool}
\subsubsection{Transcription Rate Expression Tool}

Building realistic genetic networks require realistic rate expressions for gene regulation. Such rate expressions can often be complex, and they may not be readily known. A user who is not familiar with gene regulation rate expressions may enter incorrect rate expressions resulting in an inaccurate network model. To avoid such errors as well as reduce the burden on the user, Athena provides a visual tool, the Transcription Rate Tool (shown in Figure \ref{fig:07}, that automatically formulates the transcription rate equation when the user provides the binding affinities of the transcription factors. The derived rate expression assumes that the binding and unbinding of transcription factors are at equilibrium. 

Since diffusion and binding of transcription factors are generally much faster than transcription and translation, it can often be assumed that diffusion and binding processes have reached equilibrium when modelling transcriptional processes. Under this assumption, a single transcriptional rate expression can be derived if the dissociation coefficients, $k_d$, for the transcription factors are provided. The Transcription Rate Tool provides the ability for a user to specify different $k_d$ for different states of the promoter region. For example, if two transcription factors bind co-operatively, the $k_d$ might be lower for one when the other is bound. The transcription rate tool allows deriving rate expressions for such situations by allowing a user to provide different $k_d$ values for different combinations of bound transcription factors, as shown in Figure \ref{fig:07}. 

\subsubsection{Scripting Console}

In order to allow experienced users of Athena to put together workflows to improve their productivity we integrated the IronPython \cite{ironpython} runtime in Athena. This enables advanced users to use the popular Python scripting language to program against the Athena object model. All aspects of the current model can be modified, and all plug-ins are available for invocation and modification. For example, in order to bring the current model to steady state and print all steady state concentrations, a user can write a short script that invokes the steady state function in the Systems Biology Workbench plug-in (mentioned in the next section).

% of the current model the following script could be employed: 
%
%\begin{listing}
%
%def SteadyState():
%	
%  #""" This method calculates the steady state and prints the values of
%       floating species at steady state."""
%	
%  sim.loadSBML(util.GetSBML());
%  print "\nCloseness to steady state (sum of squares): ", sim.steadyState();
%  names = sim.getFloatingSpeciesNames();
%  values = sim.getFloatingSpeciesConcentration();
%  for i in range(0, len(names)):
%  	print "    species: ", names[i], " steady state value: ", values[i]  	
%
%\end{listing}

This scripting interface can easily be retargeted to take advantage of another Dynamic Language Runtime (DLR) \cite{Hugunin} scripting language, such as IronRuby.

\subsection{Plug-ins for Model Analysis}

Since models constructed using Athena can be converted to the SBML format, all models constructed in Athena can be analyzed using any of the tools in the Systems Biology Workbench. An interface to the R statistical language is also incorporated in Athena, allowing any module in Athena to be simulated and analyzed using R. 

\subsubsection{Systems Biology Workbench}

%The script in the last example employs IronPython to bring a model to steady state and print the steady state concentrations. Underneath this workflow a resource sharing framework is used to perform the actual work. 
The Systems Biology Workbench (SBW) \cite{Hucka_SBW,Bergmann_sbw}, is both a platform and programming language resource sharing framework, as well as a suite of software tools for the analysis, creation, simulation and visualization of biochemical networks. 
Athena accesses the Systems Biology Workbench for a variety of tasks: When models without layout information are loaded, a dedicated SBW module generates a layout for it. When a SBML model is imported into Athena, a SBW module will translate the model, and read the layout information. Also simulations and parameter sweeps will be done by employing a SBW enabled simulator. This happens behind the scenes, and a user of Athena will not notice that SBW is working. 
A user of Athena can directly interact with SBW by employing the scripting console or by selecting one of the available SBW Analyzers from the SBW menu. The SBW menu allows taking a snapshot of the currently active model and send it to a variety of software modules, be it for (deterministic or stochastic) simulation, bifurcation analysis, visualization or to export the model to a variety of other languages. 

\subsubsection{R Integration}

The open source statistical language, R \cite{R}, is commonly used by biologists and bioinformaticians.  Athena converts all modules that are loaded on the main screen into R differential equation functions. The R window is integrated into Athena (not shown), allowing users to perform any analysis and plots that R can perform inside Athena.  Additional R functions, such as Gillespie simulation \cite{Gillespie} and a steady state solver, are included with Athena. With the R plug-in, it is possible to perform analysis such as time series analysis, calculate covariance matrices, matrix operations, optimization, and various other capabilities.

\subsection{Extensibility}

In order to be able to support Athena in the future, we built Athena on top of an extensible plug-in architecture. That is, apart from the rendering canvas (which displays all graphical elements of the models) and a basic selecting and moving tool, all further functionality was implemented using plug-ins. There are four plug-in interfaces: 

\begin{itemize}
\item IGlyph: A graphical object that can be part of a model: such as promoters, genes or metabolites,
\item INonVisualTool: If a tool implements this interface, it can react on certain key strokes, or mouse actions on model elements it registered to. Most of Athena's basic actions are implemented this way: (i.e.: alignment tool: which is used to align the current selection of model elements, insert tool: for adding new elements to the model or connecting model elements). Any tool that does not give visual feedback and does not require additional user input would be implemented using this interface.
\item IVisualTool: This interface possesses all features of the INonVisualTool interface, but additionally can draw annotations onto the drawing canvas before and after all model elements are drawn. An example for a tool of this category includes the Grid tool, which helps to align elements precisely onto a user specified grid.  
\item IGUITool: This interface gives a plug-in the most flexibility. It possesses the capabilities of a non-visual tool, but additionally can provide its own graphical user interface. All of Athena's advanced features are implemented this way, such as the simulation panel, the R integration (see below) or the database integration to name but a few.
During program startup Athena will load all plug-ins from the plug-in directory. Additional plug-ins can be loaded by dragging and dropping them onto the application, or by using the scripting console.
\end{itemize}

\subsection{Exchange Formats}
Exchange formats play a vital role for modeling software tools. With the increasing popularity of Systems Biology \cite{Klipp} a large number of software tools has become available to aid modelers in their task. Most of these software tools support one of the de facto model exchange standards SBML \cite{SBML} or CellML \cite{CellML}. 
In order to provide users with easy exchange of models, Athena can import and export SBML models. Via SBW modules, Athena will also import standardized SBML annotations such as the SBML Layout Extension and its successor the SBML Render Extension. Thus models generated by CellDesigner or JDesigner will be displayed correctly. 
Athena also supports the Jarnac script format \cite{Jarnac}. Falling back to the SBW SBML Translators any Athena model can also be exported into a variety of other programming languages, such as C/C++, FORTRAN, Java, Matlab, Mathematica or XPP.

\section{Discussion}

Athena captures the essential features that are required from a synthetic biology tool. Models can be constructed, simulated, and saved as modules. The database of parts allows construction of networks from real parts. Sequence information allows the network to be converted to a DNA sequence that can be synthesized. The plug-in interface allows other features to be added to Athena.

The ability to construct modular networks and build new networks by connecting existing ones permits various questions that are relevant to synthetic biology to be answered readily. For example, the preservation of a device's function is always questionable when it is used in different contexts. Athena can answer this question readily by connecting the device (which would be represented as a module in Athena) of interest to different modules and observing whether the behaviour matches the expected behaviour. Similarly, one can test different devices to see which is most suitable inside a particular model. Such tests can be performed without reconstructing any of the models or making major revisions to the network being designed. 

It is not practical to build a tool that harbours all possible analysis that an engineer might need. In order to circumvent this problem, we have made Athena extensible so that new tools can be added without editing existing code. The transcription rate derivation tool, the Gillespie \cite{Gillespie} simulator, and sequence viewing tool are examples of such plug-ins. The Systems Biology Workbench integration and R integration are examples of plug-ins that enhance the analysis capabilities of Athena. Users can write plug-ins such as these to add more functionality to the software.

As the field of synthetic biology progresses, new terminologies, concepts, and databases will emerge including novel approaches to simulation \cite{Tabasco}. It is intended that Athena will move hand-in-hand with the field. There are various plans to cope with current aspects of synthetic biology. For example, synthetic biology parts are organized into categories, and the categories grow and change over time. It is planned that Athena will provide a graphical interface for viewing this category structure and placing items of a particular category into the model. In order to provide greater analytical capabilities, Athena will incorporate all the features that can be encoded in an SBML file, such as events, which would allow various other tools to interface with Athena.

\section{Conclusion}

Ideally, with such a software such as Athena, a synthetic biologist should be able to design a conceptual model and analyze the model using various tools. Once the kinetics of the model are determined, the user should be able to request the software to find the parts in the database that would match the model. Once the parts are obtained, the user would be able to make small changes to the sequence if needed and then obtain the DNA sequence for the synthetic network. Athena can perform all of these tasks where information is available. A network can be analysed using the Systems Biology Workbench, R, and any other software that supports SBML. If needed, additional analytical plug-ins can be written without modifying existing code. Once the model is satisfactory, parts that match the topology of the network can be automatically found using the built-in database plug-in. Due to the shortage of known kinetic parameters (such as binding coefficients and transcription rates) for parts, it is not possible for Athena to find parts that would fit a model's kinetic requirements. However, once the parts are found, the DNA sequence can be obtained and modified. Ideally, when the DNA is inserted into a cell, the cell would behave as predicted from the model. There may be several reasons why this may not always be the case. Some of the key parameters may be different, hence altering the behavior of the cell. Further, when parts such as promoters or ribosome binding sites are connected to genes, Athena assumes that the part behaves independently of the location at which they are placed. If there are complications in the kinetic due to secondary structures or other spacial restrictions, then connections made in Athena may not capture the biological reality. Connecting modules can have similar concerns -- if there are details that are not captured in the kinetics, the model will be inaccurate. Such problems may be alleviated when sufficient knowledge is obtianed about how each part behaves and how they interact with one another. It is our plan to make Athena even more extensible so that it will be able to adjust to any new information that is obtained about parts, and a user will be able to design models with as much detail as he pleases, thus minimizing the gap between computational and experimental results.

\section*{Acknowledgements}

This work was funded by Microsoft's Computational Challenges in Synthetic Biology 2006 Award

\begin{center}

\newpage
\begin{figure}[!h]
\centerline{\includegraphics[scale=1.0]{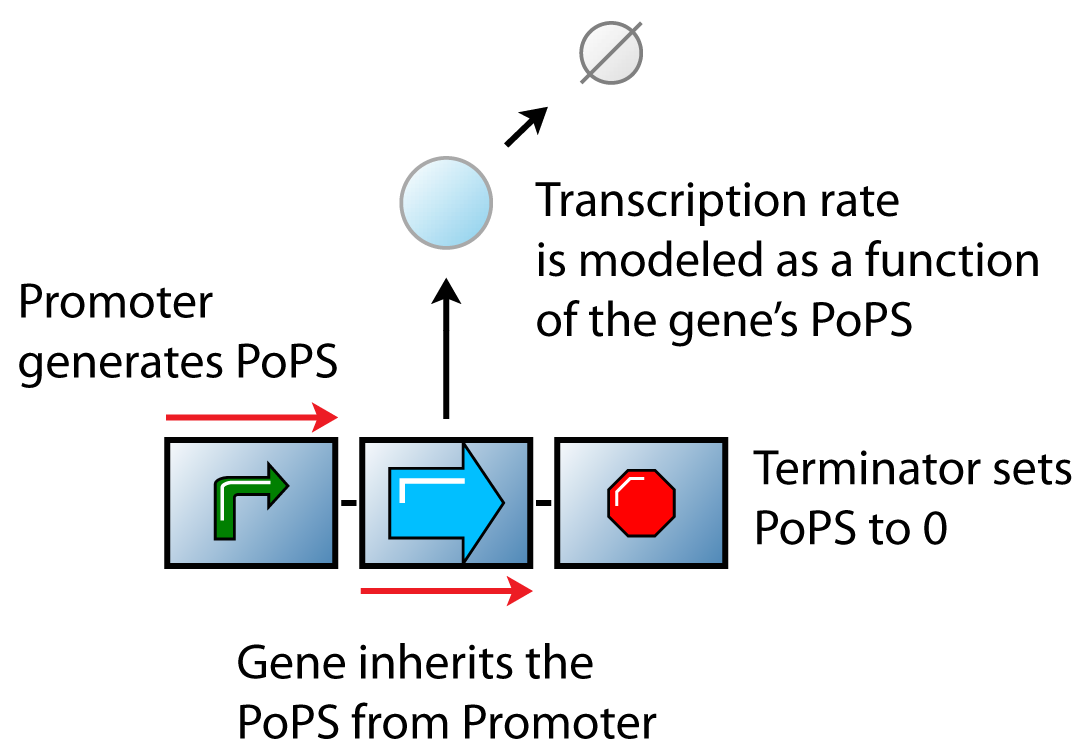}}
\caption{\noindent Regulatory networks in Athena utilize the notion of PoPS, or the rate at which RNA polymerase moves along the DNA. The promoter generates PoPS (a function of the regulating proteins), and the PoPS is carried on to the gene. The terminator, by default, stops the PoPS entirely, but the user is free to alter its PoPS output. The gene produces transcripts as a function of its PoPS. By default, the transcription rate is equal to the PoPS, but a user is free to change those rates.}\label{fig:01}
\end{figure}

\newpage
\begin{figure}[!h]
\centerline{\includegraphics[scale=0.7]{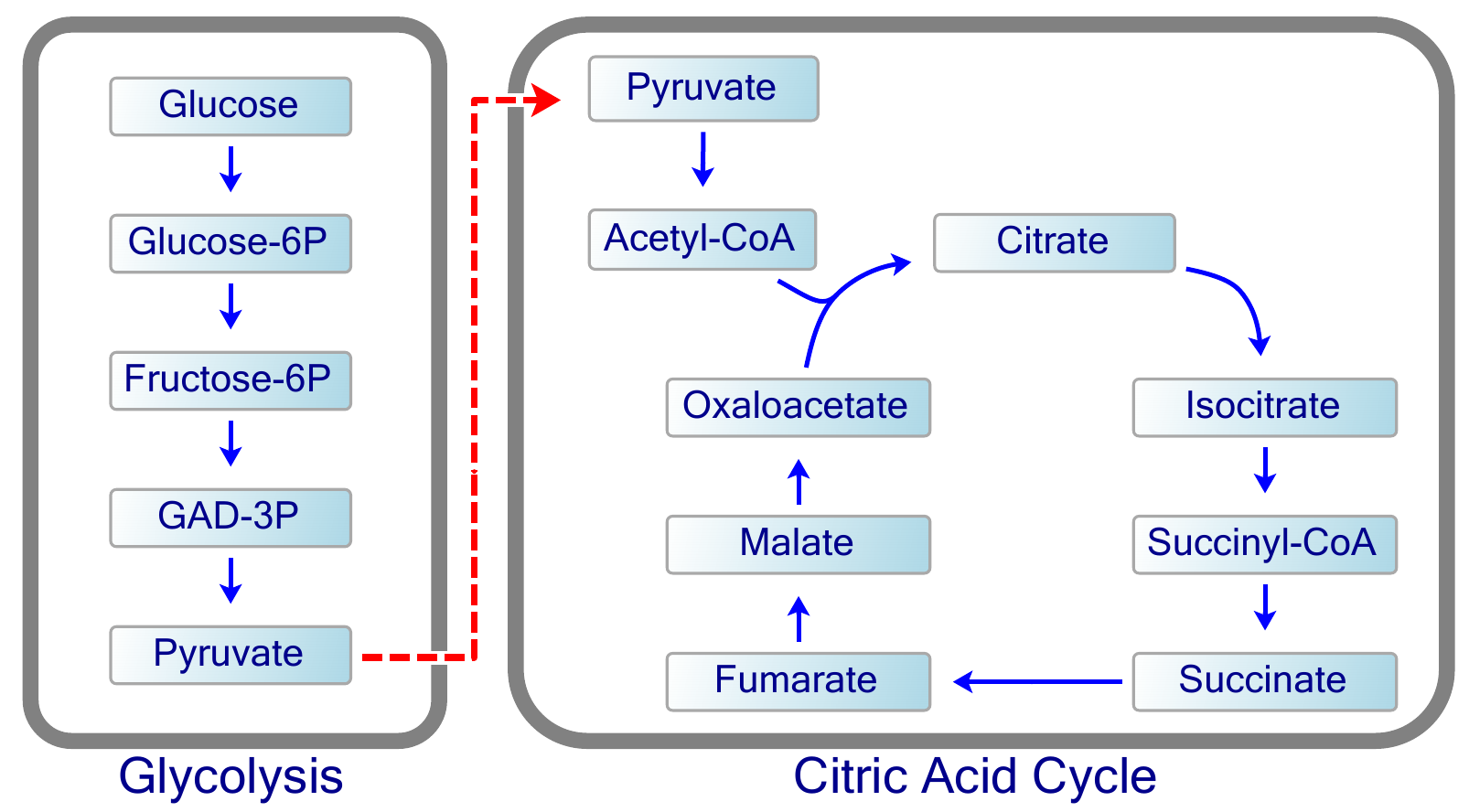}}
\caption{\noindent Two metabolic modules are connected by indicating overlapping species. The Glycolysis and the Citric Acid Cycle models are loaded into Athena and converted to modules, which are shown in this screenshot. The modules can be simulated individually and connected to one another without altering the individual modules. The dotted red connection between the two modules indicates that the Pyruvate in the Glycolysis module is the same molecule as the Pyruvate in Cytric Acid Cycle. The arrow on this connection indicates that the Pyruvate on Glycolysis model takes precedence if the names of the two molecules were different in the two modules (which is not the case in this figure). Note: the modules are simplified versions of the Glycolysis and Citric Acid Cycle in order to make the figure compact; additionally, Pyruvate is included in both modules to demonstrate the connecting feature.}\label{fig:02}
\end{figure}

\newpage
\begin{figure}[!h]
\centerline{\includegraphics[scale=0.8]{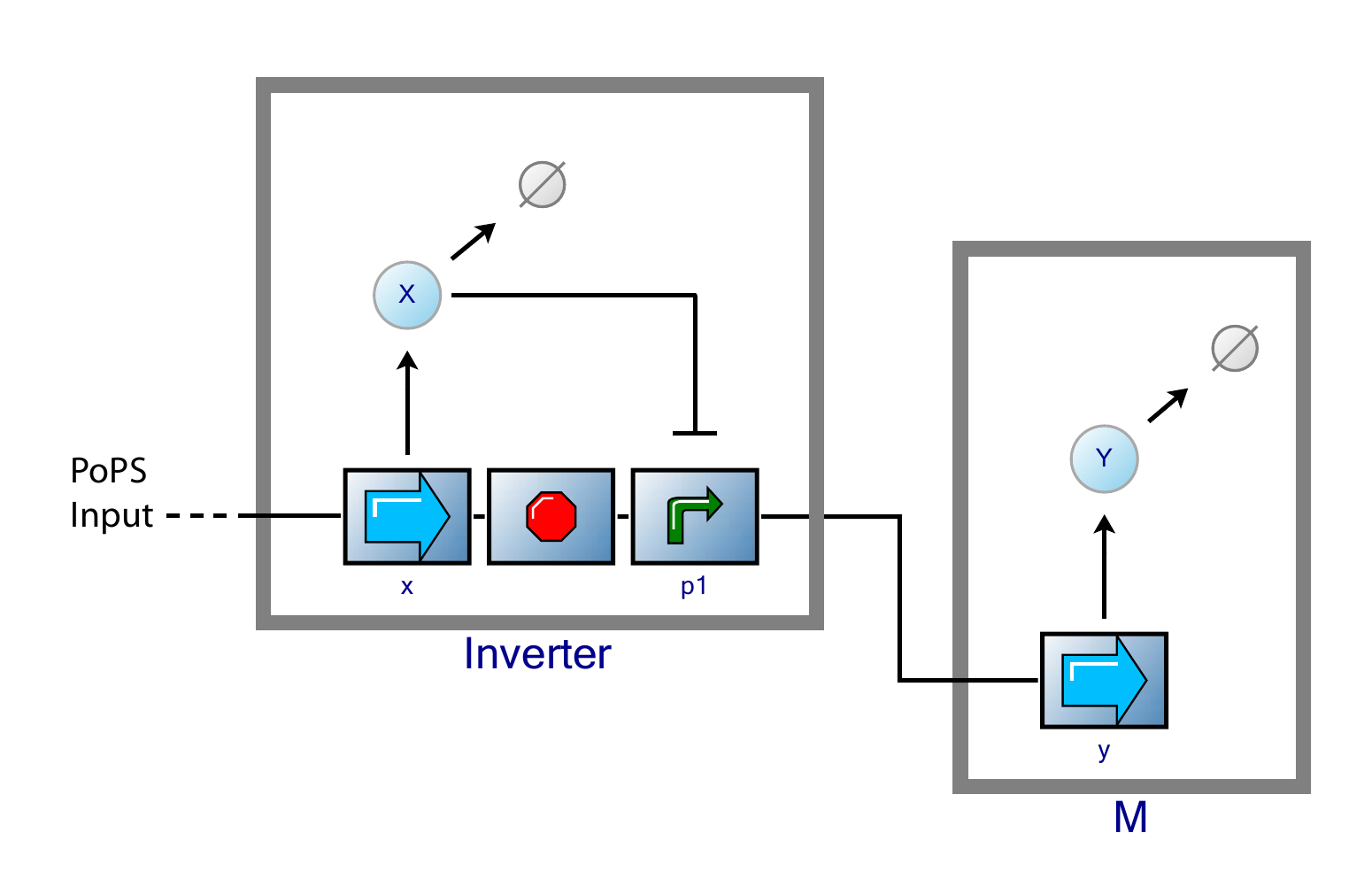}}
\caption{\noindent Connecting two modules using the PoPS interface. The Inverter module produces a low PoPS output (right of the p1 promoter) for a high PoPS input (left of gene x), and vice versa. The output PoPS can be connected to another module by connecting the promoter part of the Inverter to the input part of the other module. This feature allows PoPS devices to be constructed in Athena.}\label{fig:03}
\end{figure}

\newpage
\begin{figure}[!h]
\centerline{\includegraphics[scale=0.25]{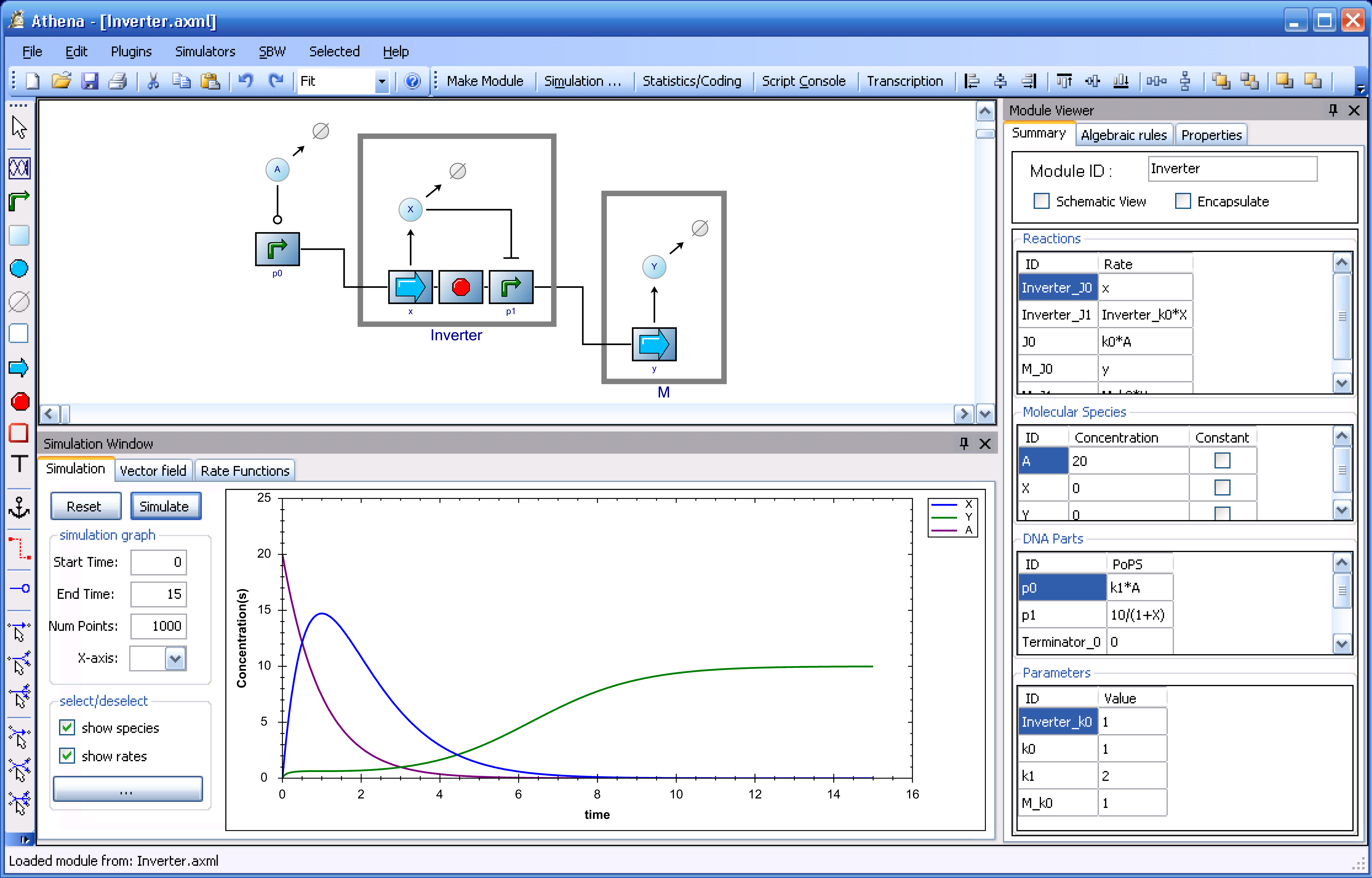}}
\caption{\noindent Screenshot of a typical Athena window. The right-hand side shows a table of all the reaction rate expressions, species concentrations, PoPS expressions, and parameters, which provides the user with a summary of the model as well as immediate control over all kinetics of the model. Selecting individual modules will allow a user to see the kinetics of the selected module. At the bottom is a simulation using the SBW simulator.}\label{fig:04}
\end{figure}

\newpage
\begin{figure}[!h]
\centerline{\includegraphics[scale=0.4]{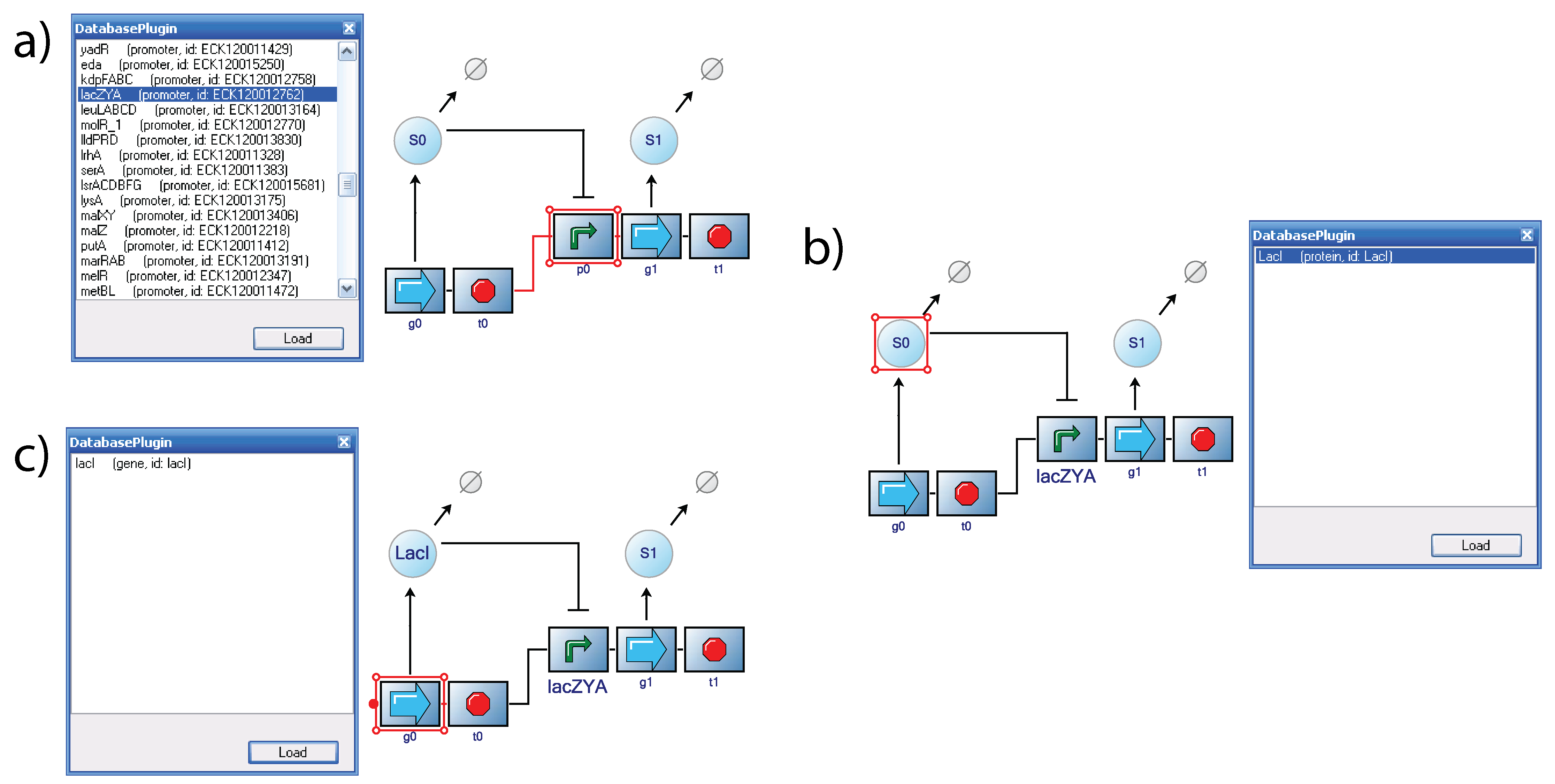}}
\caption{\noindent Using the database plug-in to find parts. (a) The promoter is replaced with one of the numerous promoters from the database. (b) Since the protein "S0" regulates a real promoter ("lacZYA" in this case), the list of available options is reduced to only the transcription factors that fit the criteria. If there are multiple proteins that meet the criteria, then the list will show all the candidates. (c) Similarly, the gene can also be substituted with the gene that produces the LacI protein. }\label{fig:05}
\end{figure}

\newpage
\begin{figure}[!h]
\centerline{\includegraphics[scale=0.5]{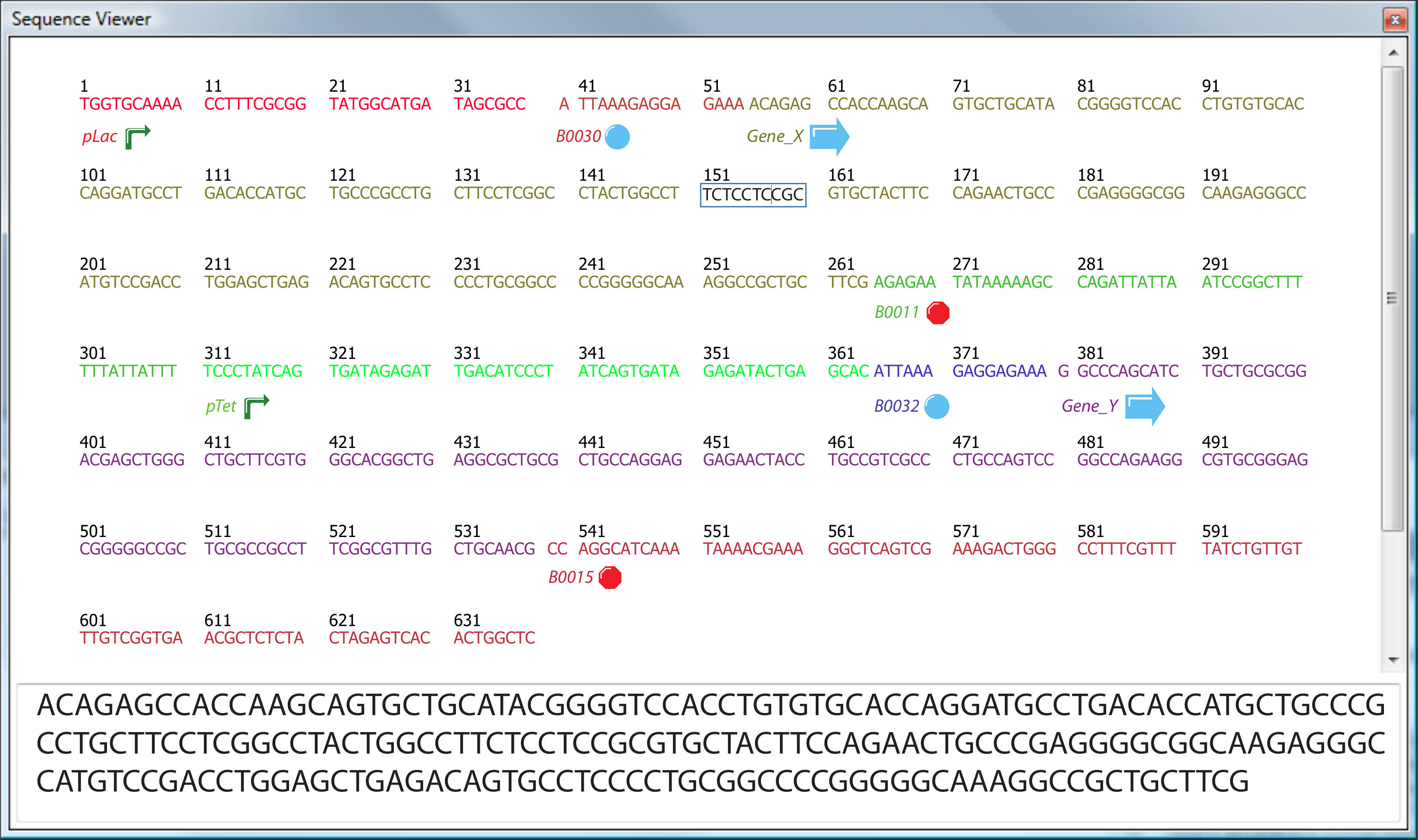}}
\caption{The Sequence Viewing Tool provides a graphical interface for editing a stretch of DNA with multiple parts. The sequences are coloured differently for the different parts, and the parts names are shown below the sequences. A text box allows for users to copy and paste other sequences.  }\label{fig:06}
\end{figure}

\newpage
\begin{figure}[!h]
\centerline{\includegraphics[scale=0.5]{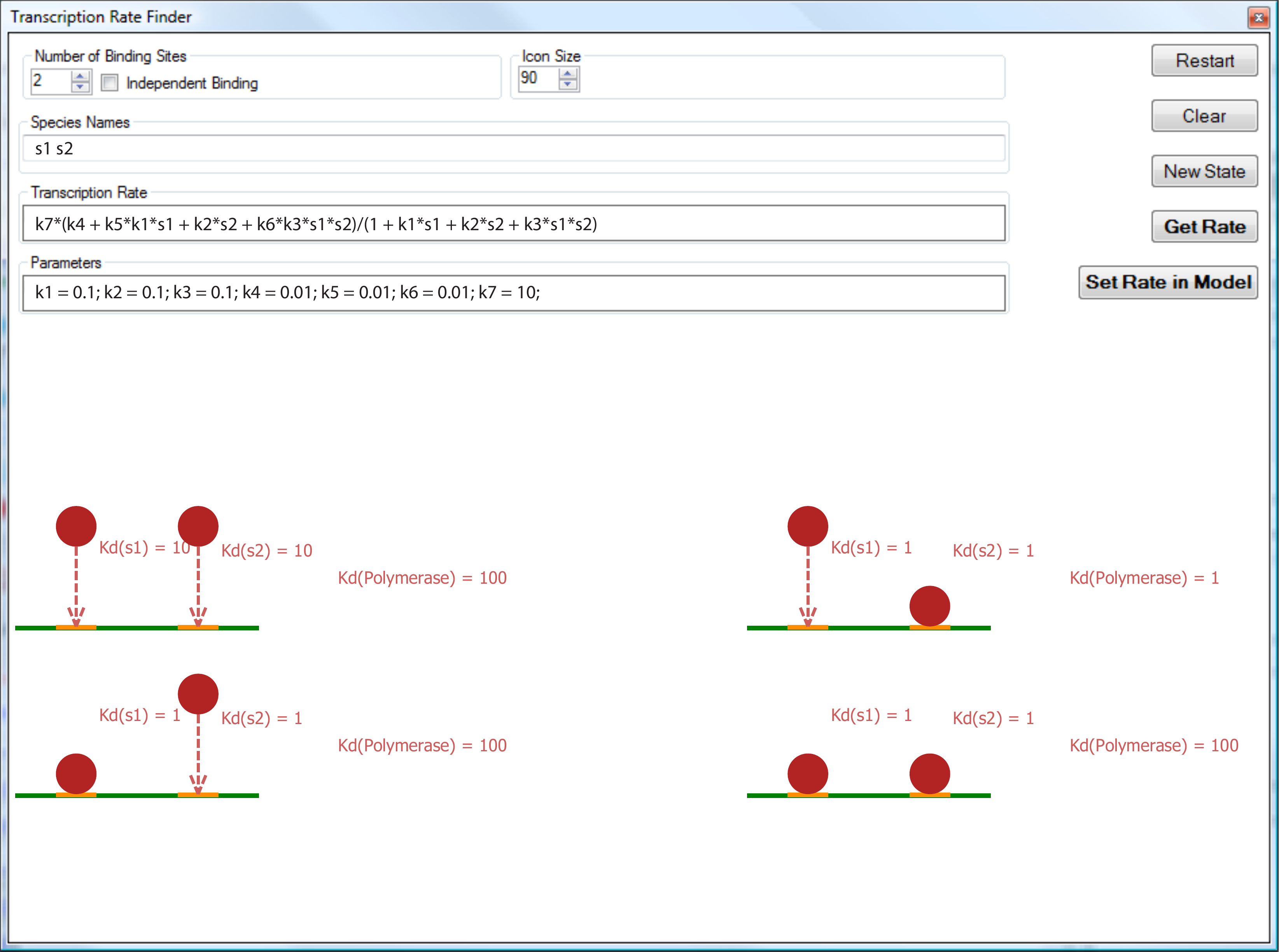}}
\caption{The Transcription Rate Tool provides the ability for users to obtain the transcription rate expression by specifying the dissociation coefficients for various transcription factors. The derived equation is based on the assumption that binding events are at equilibrium (in the time frame of transcription). The tool also checks for the Law of Detailed Balance, which says that the product of all the coefficients in a loop must be 1. }\label{fig:07}
\end{figure}

\end{center}

\end{document}